\documentclass[useAMS,usenatbib]{mn2e}

\usepackage{graphicx}

\newcommand\aj{AJ}
\newcommand\apj{ApJ}
\newcommand\mnras{MNRAS}

\newcommand{\bm}[1]{{\mbox{\boldmath $#1$}}}

\title[Scaling relations from photo-$z$s]
       {Unbiased estimates of galaxy scaling relations 
       from photometric redshift surveys}

\author[G. Rossi $\&$ R. K. Sheth]
{Graziano Rossi and Ravi K. Sheth  \\
\footnotesize Department of Physics and Astronomy, 
              University of Pennsylvania, 209 S. 33rd Street, 
              Philadelphia, PA 19104, USA\\
              Email: graziano@astro.upenn.edu, shethrk@physics.upenn.edu}

\date{Accepted ?? Received ?? ; in original form ??}

\pagerange{\pageref{firstpage}--\pageref{lastpage}}
\pubyear{2007}

\begin{document}

\maketitle

\label{firstpage}


\begin{abstract}
Many physical properties of galaxies correlate with one another, 
and these correlations are often used to constrain galaxy formation 
models.  Such correlations include the color-magnitude relation, 
the luminosity-size relation, the Fundamental Plane, etc.  
However, the transformation from observable (e.g. angular size, 
apparent brightness) to physical quantity (physical size, luminosity), 
is often distance-dependent.  
Noise in the distance estimate will lead to biased estimates of 
these correlations, thus compromising the ability of photometric 
redshift surveys to constrain galaxy formation models.  
We describe two methods which can remove this bias.  One is a 
generalization of the $V_{\rm max}$ method, and the other is a 
maximum likelihood approach.  We illustrate their effectiveness by 
studying the size-luminosity relation in a mock catalog, although 
both methods can be applied to other scaling relations as well.  
We show that if one simply uses photometric redshifts one obtains a 
biased relation; our methods correct for this bias and recover the 
true relation.  
\end{abstract}


\begin{keywords}
methods: analytical, statistical -- galaxies: formation  --- cosmology: observations.
\end{keywords}


\section{Introduction}
The `configuration space' we use to describe galaxies is large, 
but galaxies do not fill it.  
The luminosity, color, size, surface brightness, stellar velocity 
dispersion, morphology, stellar mass, star formation history and 
spectral energy distribution of a galaxy are all correlated with 
one another.  These correlations encode important information 
about galaxy formation, and so quantifying them provides important 
constraints on models.  
 
Current (e.g. SDSS, Combo-17, MUSYC, Cosmos) and planned surveys 
(e.g. DES, LSST, SNAP) go considerably deeper in multicolor photometry 
than in spectroscopy, or are entirely photometric.  For such surveys, 
reasonably accurate photometric redshift estimates are or will be 
made.  The question then arises as to which galaxy observables and 
correlations are affected by the noisy distance estimate associated 
with a photometric rather than spectroscopic redshift.  The most 
widely studied property is luminosity - clearly, errors in the distance 
result in incorrect luminosity estimates.  If not accounted for, this 
leads to a biased estimate of the luminosity function 
(e.g. Subbarao et al. 1996).  
Hence, there has been considerable effort devoted to the question of 
how to correct for this bias (e.g. Chen et al. 2003), and the problem 
has now been solved (Sheth 2007).  

The next step is to recover an unbiased estimate of not 
just the luminosity function, but the joint distribution of 
luminosity, color, size, etc., from photometric redshift datasets.    
The main goal of the present work is to provide an algorithm which 
does this for a magnitude limited photometric redshift survey.  
Because the same distance error which leads to a mis-estimate of 
the luminosity will produce a correlated mis-estimate of the size,  
we have chosen to phrase the discussion in terms of the size-luminosity 
relation - it exhibits all the features of interest.  

Section 2 illustrates the nature of the problem by showing the bias 
in the size-luminosity relation which results from treating photometric 
redshifts as though they were spectroscopic redshifts.  
This is done by constructing a mock galaxy sample and then perturbing 
the true redshifts to mimic photometric redshift errors.  
Section 3 places this problem in the more general context of inverse
problems in statistical astronomy, and argues that a deconvolution 
algorithm, such as that due to Lucy (1974), is well-suited to removing 
the bias.  
It shows the result of applying this non-parametric deconvolution 
technique to a mock galaxy sample.
Section~4 provides a maximum-likelihood formulation and solution of 
the problem.  
A final section summarizes our findings and discusses possible 
further studies and applications. 

Where necessary, we write the Hubble constant as 
$H_0 = 100h~{\rm km~s}^{-1}~{\rm Mpc}^{-1}$, and we assume a 
spatially flat cosmological model with 
$(\Omega_M,\Omega_{\Lambda}, h)=(0.3, 0.7, 0.7)$, where $\Omega_M$ and
$\Omega_{\Lambda}$ are the present-day densities of matter and
cosmological constant scaled to the critical density. 
We use $D_{\rm L}(z)$ to denote the luminosity distance; 
the angular diameter distance is $D_{\rm A}(z) = D_{\rm L}(z)/(1+z)^2$.  


\section{Correlations with observables: effect of distance errors}

\begin{figure}
\centering
 \vspace{-1.5cm}
\includegraphics[angle=0,width=0.49\textwidth]{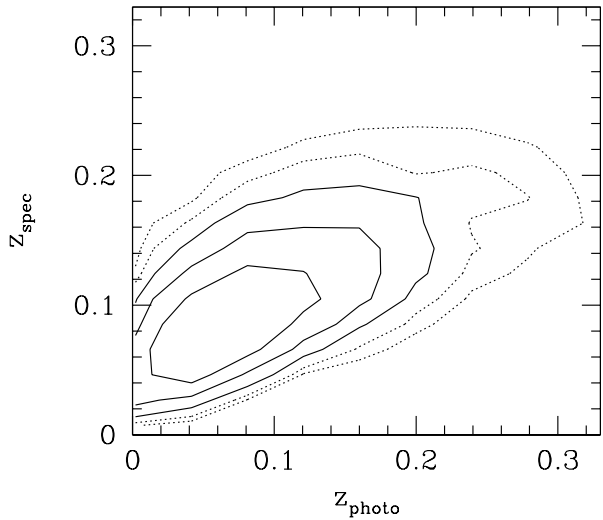}
\caption{Distribution of spectroscopic and photometric redshifts in
         our mock catalog which was set-up to mimic the SDSS early-type 
         galaxy sample.  Contours show levels which are $1/2^n$ times 
         the height of the maximum value, with $n$ running from 1 to 5.  
         The photo-$z$ error distribution was assumed to follow 
         equation (\ref{p_De_D}).}  
\label{photo_z_vs_spectro_z}
\end{figure}

In what follows, we use the luminosity-size relation to illustrate 
how photo-$z$ errors lead to biases.  

We begin by generating a magnitude-limited mock galaxy catalog with 
parameters chosen to mimic those of SDSS early-type galaxies in the 
$g$ band, following the method given by Bernardi et al. (2003). 
The redshift range is restricted to the interval $0.01 \le z \le 0.3$.
We ignored passive evolution of the luminosities and colors, as well 
as K-corrections.  
The simulated magnitude-limited catalog has a similar ${\rm d}N/{\rm d}z$
distribution to that observed, and the distribution of apparent
magnitudes, angular sizes, and velocity dispersions are very similar
to those in the real data. 

We then model photometric redshifts $\zeta$ by assuming that 
\begin{equation}
 p\big[D_{\rm L}(\zeta)|D_{\rm L}(z)\big]\,{\rm d}D_{\rm L}(\zeta) 
  = \frac{{\rm d}x}{x}\, (\gamma x)^{\gamma}\,
    \frac{\exp(-\gamma x)}{\Gamma(\gamma)}, 
 \label{p_De_D}
\end{equation}
where $x=D_{\rm L}(\zeta)/D_{\rm L}(z)$ is the ratio of the photo-$z$ 
based luminosity distance to the true one, and $\gamma = 5$.  
This distribution has $\langle x \rangle = 1$ and 
$\sigma_{x}^{2} = 1/\gamma$.  With $\gamma = 5$, this error distribution 
is substantially worse than typical photometric redshift errors.  
Figure~\ref{photo_z_vs_spectro_z} compares $\zeta$ and $z$.  
Note that the analysis which follows is not tied to this functional 
form for the photo-$z$ error distribution; we are simply using it 
to illustrate our methods.  

\begin{figure}
\centering
\includegraphics[angle=0,width=0.58\textwidth]{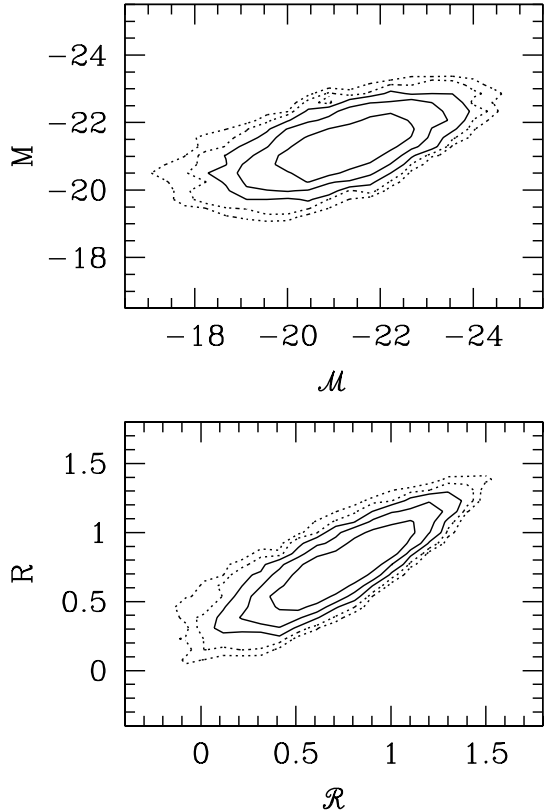}
\caption{Distributions of intrinsic and estimated absolute magnitudes
  (top panel) and sizes (bottom panel) which result from the differences 
  between true and photo-$z$ shown in the previous Figure.  }
\label{M_Me_R_Re_early_types}
\end{figure}

In what follows, we use $M$ to denote the true absolute magnitude, 
and ${\cal M}$ to denote that estimated using $\zeta$ rather than 
$z$.  And, with some abuse of notation, we use $R$ to denote 
$\log_{10}$ of the physical size $\theta D_{\rm A}(z)$, where $\theta$ 
is the measured angular size and $D_{\rm A}(z)$ is the angular diameter 
distance defined earlier.  The estimated size based on the photometric 
redshift $\zeta$ is 
\begin{eqnarray}
 {\cal R} &\equiv& \log_{10} [\theta D_{\rm A}(\zeta)] 
     = R + \log_{10} [D_{\rm A}(\zeta)/D_{\rm A}(z)]\nonumber\\
     &=& R - ({\cal M} - M)/5 - 2\log_{10} [(1+\zeta)/(1+z)];
\end{eqnarray}
in principle, there are also evolution and K-correction terms 
which we set to zero.  
Figure~\ref{M_Me_R_Re_early_types} compares ${\cal M}$ with $M$ and 
${\cal R}$ with $R$.  

\begin{figure*}
\begin{center}
\includegraphics[angle=0,width=0.8\textwidth]{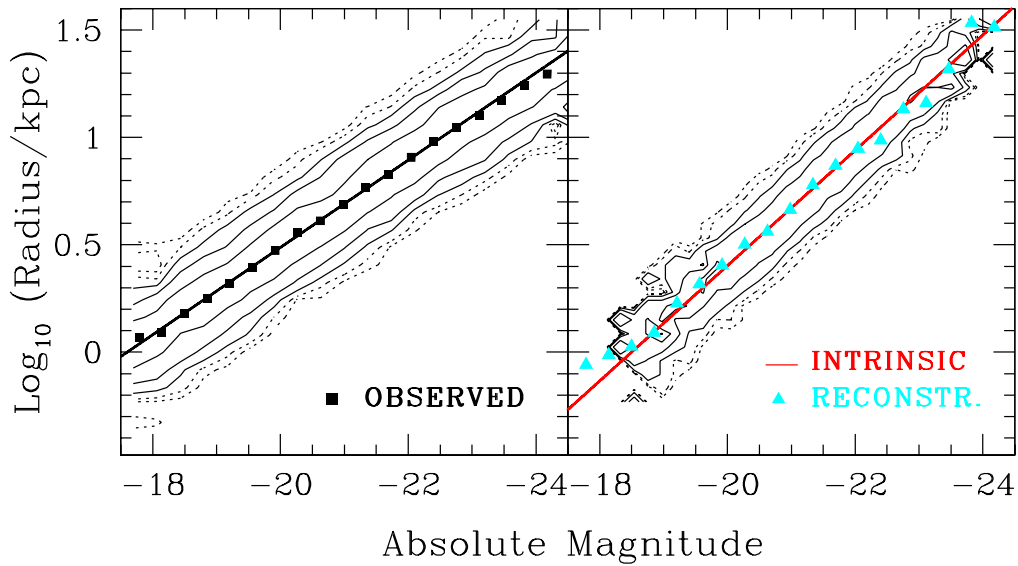}
\caption{Effect of photo-$z$ on the size-luminosity correlation in 
  our mock catalog.  In the left panel, contours and solid line 
  show the ${\cal R}-{\cal M}$ relation associated with photo-$z$s, 
  whereas the right panel shows the intrinsic $R-M$ relation.  
  Note the strong bias (shallower slope in panel on left) which 
  results from the fact that the photo-$z$ distance error moves 
  points down and left or up and right on this plot.  
  Squares in left panel show the binned starting guess for the 
  2d deconvolution algorithm, and triangles in right panel show 
  the result after 20 iterations (equation~\ref{average_eta_given_xi}). 
  Convergence to the correct solution is clearly seen.}
\label{M_R_correlation_rec}
\end{center}
\end{figure*}

The qualitative nature of the distributions in the two panels are 
easy to understand.  The distribution in ${\cal M}$ is broader 
than that in $M$, as is the distribution of ${\cal R}$ compared 
to $R$:  photometric redshift errors have broadened both distributions.
However, the changes to the estimated absolute magnitude and size are 
not independent.  Assuming an object is closer than it really is makes 
one infer a fainter luminosity and smaller size than it really has. 
These correlated changes can have a dramatic effect on the 
size-luminosity relation, since photo-$z$ errors move each galaxy 
in the $R-M$ plane left and down or right and up.  
In general, these motions are not parallel to the principal axis 
of the true relation, so the mean relation in the photo-$z$ catalog, 
$\langle {\cal R}|{\cal M} \rangle$, need not be the same as the true 
relation $\langle R|M \rangle$.  In our mock catalog,
 $\langle {\cal R}|{\cal M} \rangle \propto -0.20\,{\cal M}$, whereas 
 $\langle R|M \rangle \propto -0.27\,M$ (see Figure~\ref{M_R_correlation_rec}).


\section{A non-parametric deconvolution-like method}\label{Vmax}
Sheth (2007) shows that to extend Schmidt's (1968) $V_{\rm max}$ 
estimator of the luminosity function $\phi(M)$ so that it produces 
unbiased results in photo-$z$ surveys, it is helpful to think of 
${\cal N}({\cal M})$, the number of observed objects with estimated 
${\cal M}$, as being a convolution of the true number which have $M$, 
$N(M)$, with the probability that an object with magnitude $M$ is 
thought to have magnitude ${\cal M}$.  
The luminosity function is then estimated by first deconvolving 
the distribution of ${\cal N}({\cal M})$ to obtain $N(M)$, 
and then using the fact that $N(M) \equiv \phi(M)\,V_{\rm max}(M)$.  

For the same reason, the present problem may be thought of as a 
two dimensional deconvolution problem (or an $n$-dimensional 
deconvolution if we were interested in the full manifold, rather 
than just two-dimensional projections of it).  
Specifically, let
\begin{equation}
 N(M,R) = N(M)\,p(R|M) = V_{\rm max}(M)\,\phi(M)\,p(R|M)
\end{equation} 
denote the (true) number of galaxies with absolute magnitude $M$ 
and size $R$ in a magnitude-limited catalog.  Here $p(R|M)$ is 
the probability of having size $R$ when the magnitude is $M$.
Similarly, set 
${\cal N}({\cal M},{\cal R}) = {\cal N}({\cal M})~p({\cal R}|{\cal M})$.
Then
\begin{eqnarray}
 {\cal N}({\cal M},{\cal R}) &=& 
  \int {\rm d}M \int {\rm d}R~N(M,R)~p({\cal M},{\cal R}|M,R) \nonumber \\
  &=& \int {\rm d}M ~N[M,{\cal R}-F(M,{\cal M})] ~p({\cal M}|M),
 \label{Ne_2d}
\end{eqnarray}
where $F=({\cal M}-M)/5+2\log[(1+\zeta)/(1+z)]$. 
Our problem is to obtain a reliable estimate of the intrinsic 
$N(M,R)$ given the photo-$z$ biased ${\cal N}({\cal M},{\cal R})$ and the 
error distribution $p({\cal M}|M)$.  We do this using the deconvolution 
algorithm proposed by Lucy (1974).

Before we present our algorithm and results, it is worth noting that 
we could have attempted to invert equation~(\ref{Ne_2d}) in other ways.  
Classical naive `exact' inversion methods include matrix-quadrature
techniques (reduction of the integral equation to a linear matrix
system), polynomial expansion methods, singular function expansion and
product integration methods. 
These typically run into difficulties because the measured data 
function usually cannot provide sufficient information on the high 
frequency components of the solution. 
A standard non-classical technique (Phillips 1962; Twomey 1963;
Tikhonov 1963) is the method of regularisation. A ``regularisation''
parameter is introduced, which balances the size of the residual 
against the smoothness of the solution, and the problem is turned 
into one of minimization.  
However, there is no general strategy for choosing the optimum 
regularisation parameter; this led Lucy (1974) to formulate 
his algorithm, and is what has led us to choose his algorithm 
over these others.  One might argue that we are likely to know a 
fair amount about the expected form of the intrinsic distribution 
(e.g., luminosity functions are rather smooth, and conditional 
distributions tend to be bell-shaped), so it may be that these 
other methods are worth pursuing further.  This is the subject 
of work in progress.  


\subsection{The deconvolution algorithm}
The general two-dimensional problem is that of estimating the 
frequency distribution $\Psi (\xi', \eta')$ of the quantities 
$\xi'$ and $\eta'$ when the available measures 
${x'_1, y'_1; x'_2, y'_2; ..., x'_N, y'_N}$ are a finite sample 
drawn from an infinite population characterized by:
\begin{equation}
 \Phi(x,y) = \int {\rm d}\xi \int {\rm d}\eta~\Psi(\xi, \eta)~p(x, y|\xi, \eta).
 \label{Phi_2d}
\end{equation}
Here $\Phi(x,y)$ is the function accessible to measurement  
and $p(x, y|\xi, \eta)$ is the conditional probability that $x'$
will fall in $[x,x+{\rm d}x]$ when it is known that $\xi' \equiv \xi$, 
and that $y'$ will fall in $[y,y+{\rm d}y]$ when $\eta' \equiv \eta$. 
In many cases $\Phi$ and $\Psi$ represent probability density
functions, so they obey normalization and non-negativity constraints.

The iterative procedure for generating estimates to $\Psi$ presented
in Lucy (1974) is 
\begin{equation}
 \Psi^{r+1} (\xi, \eta) = \Psi^{r}(\xi, \eta)
  \int {\rm d}x \int {\rm d}y~\frac{\tilde{\Phi} (x,y)}{\Phi^r (x,y)}~p(x,y|\xi,\eta),
 \label{Psi_2d}
\end{equation}
where 
\begin{equation}
 \Phi^r(x,y) = \int {\rm d}\xi \int {\rm d}\eta~\Psi^r(\xi, \eta)~p(x,y|\xi,\eta).
 \label{Phi_r}
\end{equation}
The index $r$ indicates the $r$th iteration in the sequence of 
estimates, and $\tilde \Phi$ is an approximation to $\Phi$ 
obtained from the observed sample.
Convergence is achieved if $\Phi^r = \tilde{\Phi}$. The starting
approximation $\Psi^{0}(\xi,\eta)$ should be a smooth, non-negative
function having the same integrated density as the observed distribution. 
The extension to $n$-dimensions is obvious, although, if the 
number of dimensions is large, then performing the multi-dimensional 
integrations efficiently may become challenging.  

The two-dimensional problem simplifies if, as happens in our problem, 
\begin{eqnarray}
 p(x,y|\xi,\eta) &=& p(x|\xi)\, p(y|\xi,\eta,x) \nonumber\\
                 &=& p(x|\xi)\, \delta_{\rm D}\Bigl[y = \eta - F(x,\xi)\Bigr]
\end{eqnarray} 
because the delta function simplifies one of the integrals.  
The iterative scheme becomes
\begin{equation}
 \Psi^{r+1} (\xi, \eta) = \Psi^{r}(\xi, \eta) 
  \int {\rm d}x~\frac{\tilde{\Phi} [x, \eta-F(x,\xi)]}
               {\Phi^r [x, \eta - F(x,\xi)]}~p(x|\xi)
\label{Psi_2d_simpler}
\end{equation}
with
\begin{equation}
 \Phi^r(x,y) = \int {\rm d}\xi~ \Psi^r[\xi, y + F(x,\xi)]~p(x|\xi).
\end{equation}
Partial distributions can be easily computed via marginalization 
from $\Psi^{r+1}$:
\begin{equation}
 \Psi^{r+1}(\xi) = \int {\rm d}\eta~\Psi^{r+1}(\xi,\eta),
 \label{Psi_2d_xi}
\end{equation}
\begin{equation}
 \Psi^{r+1}(\eta) = \int {\rm d}\xi~\Psi^{r+1}(\xi,\eta).
 \label{Psi_2d_eta}
\end{equation}
and 
\begin{equation}
 \langle \eta|\xi \rangle^{r+1} = \int {\rm d}\eta~\eta~
                    \frac{\Psi^{r+1}(\eta,\xi)}{\Psi^{r+1}(\xi)}.
 \label{average_eta_given_xi}
\end{equation}
The generalization to $n$-dimensions is a straightforward 
extension of the expressions above, so we do not present explicit 
expressions.  Note that, just as happens in the 2-dimensional 
case presented here, delta-functions will reduce the $n$-dimensional  
problem to a simple one-dimensional integral, because the same 
distance error affects all $n$ quantities.


\begin{figure}
 \vspace{-1cm}
\begin{center}
\includegraphics[angle=0,width=0.5\textwidth]{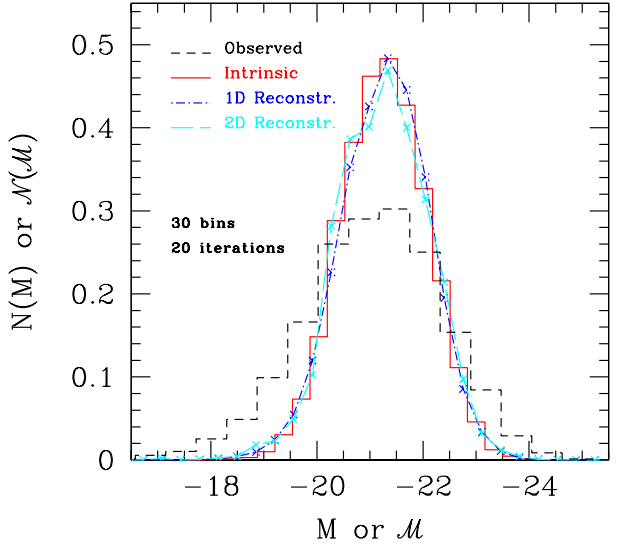}
\caption{Reconstruction of the intrinsic $N(M)$ distribution from the
  distribution of estimated redshifts. Dashed histogram shows the
  observed absolute magnitude distribution, used as a starting
  guess. Jagged lines show the reconstructed intrinsic distribution
  after 20 iteration, using the simpler 1d algorithm (blue) or a 2d
  iterative scheme (cyan).}
\label{N_M_rec}
\end{center}
\end{figure}

\begin{figure}
 \vspace{-1.5cm}
\begin{center}
\includegraphics[angle=0,width=0.5\textwidth]{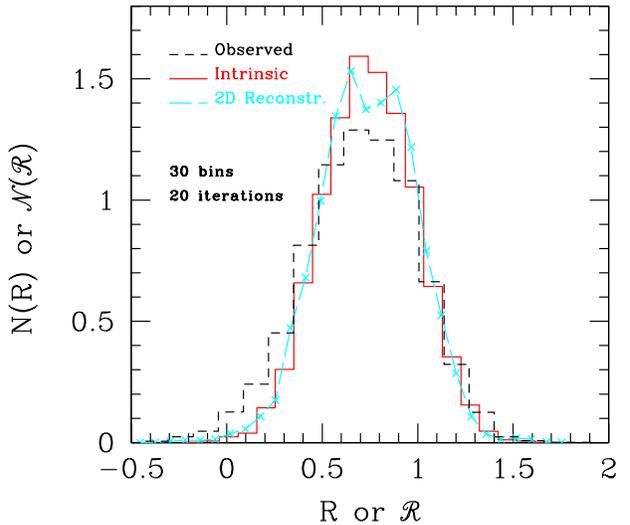}
\caption{Reconstruction of the intrinsic $N(R)$ distribution from the
  distribution of estimated redshifts.  Line styles same as previous 
  Figure.  }
\label{N_R_rec}
\end{center}
\end{figure}

\begin{figure*}
 \vspace{-0.5cm}
\begin{center}
\includegraphics[angle=0,width=0.9\textwidth]{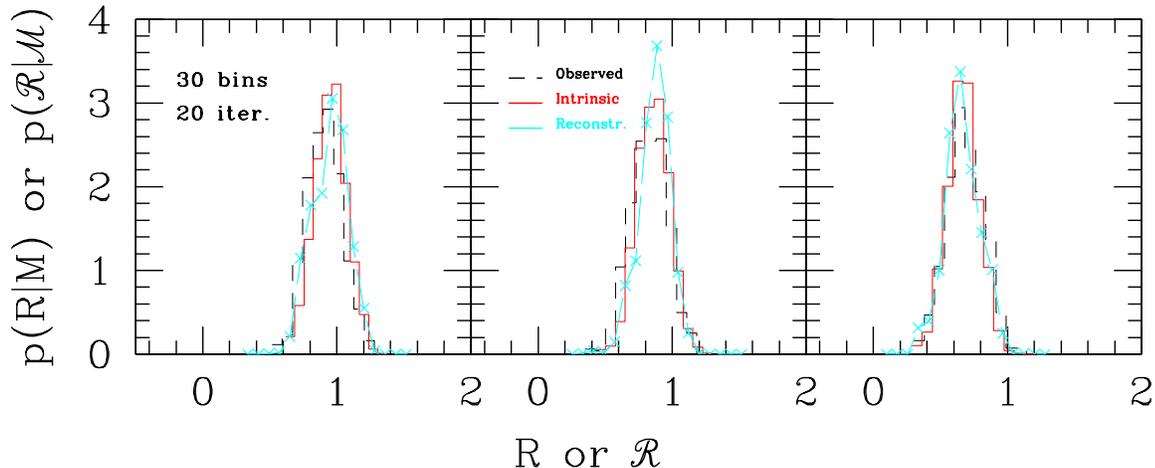}
\caption{Examples of reconstructed conditional distributions $p(R|M)$ 
  bins in magnitude of width $\Delta M=0.355$ centred on 
  $M=-22.046, -21.691, -20.982$.
  Jagged lines show the distributions recovered by the 2d deconvolution
  algorithm after 20 iterations.}
\label{p_R_given_M_rec}
\end{center}
\end{figure*}

\subsection{Results}
The formalism outlined above is readily applicable to the 
size-luminosity correlation if we interpret ($x,y$) as the estimated 
absolute magnitudes and sizes (${\cal M},{\cal R}$), and ($\xi,\eta$) 
as the true intrinsic ones, ($M,R$).

Figures~\ref{N_M_rec} and~\ref{N_R_rec} show how well this method 
recovers the intrinsic distribution of absolute magnitudes and 
sizes (solid histograms).  The broad dashed histograms show the 
photo-$z$ derived ${\cal M}$ and ${\cal R}$ distributions.  These were used 
as a convenient starting guess in the deconvolution algorithm, 
although prior knowledge about the expected intrinsic shapes could 
have been used instead.  The reconstruction after 20 iterations is 
shown by the jagged lines.
Of course, we could have chosen to reconstruct $N(M)$ directly from 
${\cal N}({\cal M})$, using the 1d deconvolution algorithm outlined in Sheth (2007).  
This procedure converges in about 5 iterations to the distribution 
shown by the crosses.  
Note how well the deconvolved distributions match the intrinsic ones.  

A more stringent test is to check if the conditional distributions, 
$p(R|M)$, are also well recovered.  
Figure~\ref{p_R_given_M_rec} shows $p(R|M)$ for three bins of width 
$\Delta M=0.355$ centred at $-22.046$, $-21.691$ and $-20.982$.  
Clearly, the method works well.  

The means of these recovered distributions can be used as an estimate 
of the recovered size-luminosity relation:  $\langle R|M\rangle$.  
Recall we had noted that this relation was rather strongly biased 
because use of the photo-$z$ distance estimate means that the error 
in the size is correlated with that on the luminosity.  
The squares in the left hand panel of Figure \ref{M_R_correlation_rec} 
show the starting guess for our algorithm, and the triangles in the 
right hand panel show the reconstructed relation obtained from the 
deconvolution procedure --- it is in excellent agreement with the true 
one.


\section{A maximum-likelihood method}\label{ml}
Sheth (2007) describes an algorithm which produces a maximum-likelihood 
estimate of the luminosity function from magnitude limited photo-$z$ 
datasets.  It is straightforward to extend that analysis to the present 
case, in which the quantity of interest is not just the distribution 
of luminosities, but the joint distribution of luminosity and other 
observed physical parameters.  

Let ${\bm {\cal M}}_i$ denote the vector of physical quantities 
for galaxy $i$ estimated using the photometric redshift $\zeta_i$ when 
computing distances, and let 
 ${\cal N}({\bm {\cal M}}_i, \zeta_i|{\bm a})$
denote the number of galaxies in a magnitude limited catalog 
that have estimated redshifts $\zeta_i$ and estimated luminosities, 
sizes, etc., ${\bm{\cal M}}_i$, when the model for the 
intrinsic joint distribution of physical quantities is specified 
by the parameters ${\bm a}$.  
Then, dropping the understood $i$-index dependence,
\begin{equation}
 {\cal N}({\bm {\cal M}},\zeta|{\bm a}) = 
  \int {\rm d}z {{\rm d}V_{\rm c}\over {\rm d}z}\,
   \phi({\bm M}|{\bm a})\, p\Bigl(\zeta|z,m({\bm {\cal M}},\zeta)\Bigr),
\end{equation}
where $p(\zeta|z,m)$ is the photo-$z$ redshift error distribution, 
and $\phi({\bm M}|{\bm a})$ denotes the true joint distribution of 
physical quantities, but evaluated at values which account for the 
fact that the same photo-$z$ error which affects the absolute 
magnitudes affects the other observables.  
E.g., in the size-luminosity relation we have been considering, 
\begin{equation}
 {\bm {\cal M}} = ({\cal M},{\cal R}) \qquad {\rm and}\qquad {\bm M} = (M,R),
\end{equation}
where 
\begin{eqnarray}
 \label{Meff}
 M &=& {\cal M} - 5 \log_{10} [D_{\rm L}(z)/D_{\rm L}(\zeta)]\\
 \label{Reff}
 R 
   &=& {\cal R} + \log_{10} [D_{\rm A}(z)/D_{\rm A}(\zeta)].
\end{eqnarray}
The factor of $m$ is the apparent magnitude associated with 
photo-$z$ redshift $\zeta$ and absolute magnitude ${\cal M}$; 
of course, $m$ is the same for true redshift $z$ and absolute 
magnitude $M$.  

The predicted number of objects with photometric redshift $\zeta$ 
when the model has parameters ${\bm a}$ is 
\begin{equation}
 {\cal N}(\zeta|{\bm a}) = 
 \int {\rm d}{\bm {\cal M}}\, {\cal N}({\bm {\cal M}},\zeta|{\bm a}).
\end{equation}
If the redshift-error distribution is independent of $m$, then 
\begin{eqnarray}
 {\cal N}(\zeta|{\bm a}) &=& \int {\rm d}z\,({\rm d}V_{\rm c}/{\rm d}z)\,
                     S(z,{\bm a})\,p(\zeta|z) \nonumber\\
                  &\equiv& \int {\rm d}z\,N(z|{\bm a})\,p(\zeta|z),  
\end{eqnarray}
where 
\begin{equation}
 S(z,{\bm a}) = \int_{M_{\rm max}(z)}^{M_{\rm min}(z)} {\rm d}M\,\phi(M).  
\end{equation}
This shows that ${\cal N}(\zeta|{\bm a})$ is just the convolution of 
the intrinsic redshift distribution (in a flux-limited catalog) with 
the redshift-error distribution.  

The expressions above generalize those given in Sheth (2007), 
where ${\bm {\cal M}} = {\cal M}$ and ${\bm M} = M$, with $M$ given 
by equation~(\ref{Meff}).  
Hence, by analogy to when the distances are known accurately, the 
likelihood to be maximized is (reintroducing the index $i$)
\begin{equation}
 {\cal L}({\bm a}) = \prod_i\, p_i, \qquad {\rm where} \qquad 
      p_i = {{\cal N}({\bm {\cal M}}_i, \zeta_i|{\bm a})\over 
             {\cal N}(\zeta_i|{\bm a})}.
 \label{mlphotoz}
\end{equation}
The analysis in Sheth (2007) can now be followed to show analytically 
that this is indeed the appropriate expression for the likelihood, so 
we do not reproduce it here.


\section{Discussion and future work}
We presented two algorithms for reconstructing the intrinsic
correlations between distance-dependent quantities in apparent 
magnitude limited photometric redshift datasets.  One was a 
generalization of the non-parametric $V_{\rm max}$ method 
(Section~\ref{Vmax}), and the other used a maximum-likelihood 
approach (Section~\ref{ml}).  

Both our reconstruction methods assume that the distribution of 
photo-$z$ errors is known accurately.  In practice, this means that 
spectroscopic redshifts are available for a subset of the data.  
The question then arises as to whether or not the number of spectra 
which must be taken to specify the error distribution reliably is 
sufficient to also provide a reliable (spectroscopic!) estimate of 
these scaling relations.  If so, what is the basis for deciding that 
it is worth reconstructing these relations from the photo-$z$ data?  
This is the subject of work in progress, although the methods presented 
in this paper assume that such reconstructions will indeed be necessary.  
E.g., if the spectra are not simply a random subset of the magnitude 
limited photometric sample, then it may be difficult to quantify 
and so correct for the selection effects associated with the 
spectroscopic subset.  

We used the size-luminosity relation in a mock catalog which had 
realistic choices for the correlation to illustrate the biases 
which are present and must be corrected if photometric redshift 
datasets are to provide reliable estimates of galaxy scaling relations
(Figures~\ref{M_Me_R_Re_early_types} and~\ref{M_R_correlation_rec}).  
We showed that our iterative deconvolution scheme provides a simple 
and reliable correction of this bias 
(Figures~\ref{M_R_correlation_rec}--\ref{p_R_given_M_rec}).  Note 
that although we have illustrated our methods using a 2-dimensional 
distribution, the extension to $n$-correlated variables is trivial.  

Because our algorithm permits the accurate measurement of many 
scaling relations for which spectra were previously thought to be 
necessary (e.g. the color-magnitude relation, the size-surface 
brightness relation, the Photometric Fundamental Plane), we hope that 
our work will permit photometric redshift surveys to provide more 
stringent constraints on galaxy formation models at a fraction of 
the cost of spectroscopic surveys.  

Our results may have other applications.  For example, 
Bernardi (2007) has highlighted a bias associated with the correlation 
between stellar velocity dispersion $\sigma$ and luminosity $L$ which 
arises if the distance indicator used to estimate $L$ is correlated 
with $\sigma$ (as may happen in the local Universe, where peculiar 
velocities make spectroscopic redshifts unreliable distance estimators).  
It may be that the methods presented here would allow an accurate 
reconstruction of the true relation from the biased one.  
This is the subject of on-going work.


\section*{Acknowledgments}
This work was supported by NSF 0520677.  
RKS thanks the Aspen Center for Physics for hospitality during the 
Summer of 2007, during which time he had interesting conversations 
with N. Padmanabhan about the pros and cons (but mainly the pros!) 
of regularized inversion techniques.  



\label{lastpage}

\end{document}